# IDEAS: Immersive Dome Experiences for Accelerating Science

ASTRO 2020 White Paper

State of the Profession

Jacqueline K. Faherty (American Museum of Natural History, Hayden Planetarium, jfaherty@amnh.org), Mark SubbaRao (Adler Planetarium, International Planetarium Society msubbarao@adlerplanetarium.org), Ryan Wyatt (California Academy of Sciences, rwyatt@calacademy.org ), Anders Ynnerman (Linköping University), Neil deGrasse Tyson (American Museum of Natural History, Hayden Planetarium), Aaron Geller (Adler Planetarium, Northwestern University), Maria Weber (Adler Planetarium, University of Chicago and soon-to-be planetarium director at Delta State University), Philip Rosenfield (AAS WorldWide Telescope), Wolfgang Steffen (Instituto de Astronomía, UNAM), Gabriel Stoeckle (Natural History Museum Vienna, gabriel.stoeckle@nhm-wien.ac.at), Daniel Weiskopf (Visualization Research Center, University of Stuttgart, weiskopf@visus.uni-stuttgart.de), Marcus Magnor (Computer Graphics Lab, TU Braunschweig), Peter K. G. Williams (Center for Astrophysics | Harvard & Smithsonian, American Astronomical Society), Brian Abbott (American Museum of Natural History, Hayden Planetarium, abbott@amnh.org), Lucia Marchetti (University of Cape Town, Iziko Planetarium & Digital Dome, marchetti.lu@gmail.com), Thomas Jarrrett (University of Cape Town, Iziko Planetarium & Digital Dome), Jonathan Fay (AAS WorldWide Telescope), Joshua Peek (jegpeek@stsci.edu Space Telescope Science Institute & Johns Hopkins University), Or Graur (Center for Astrophysics, Harvard & Smithsonian), Patrick Durrell (Youngstown State University Ward Beecher Planetarium prdurrell@ysu.edu), Derek Homeier (Förderkreis Planetarium Göttingen, dhomeie@gwdg.de), Heather Preston (Calusa Nature Center & Planetarium heather@calusanature.org), Thomas Müller (Haus der Astronomie & MPIA, tmueller@mpia.de), Johanna M Vos (American Museum of Natural History, jvos@amnh.org), David Brown (Microsoft Research, dabrown@microsoft.com), Paige Giorla Godfrey (Slooh, paige@slooh.com), Emily Rice (CUNY Macaulay Honors College, erice@amnh.org), Daniella Bardalez Gagliuffi (American Museum of Natural History, dbardalezgagliuffi@amnh.org),  Alexander Bock (Scientific Computing and Imaging Institute, University of Utah, mail@alexanderbock.eu), James Hedberg (City College of New York, jhedberg@ccny.cuny.edu), Drew Rosen (Stony Brook University, drew.rosen@stonybrook.edu), Carter Emmart (American Museum of Natural History, carter@amnh.org)

# EXECUTIVE SUMMARY:

Astrophysics lies at the crossroads of big datasets (such as the Large Synoptic Survey Telescope and Gaia), open source software to visualize and interpret high dimensional datasets (such as Glue, WorldWide Telescope, and OpenSpace), and uniquely skilled software engineers who bridge data science and research fields. At the same time, more than 4,000 planetariums across the globe immerse millions of visitors in scientific data. We have identified the potential for critical synergy across data, software, hardware, locations, and content that—if prioritized over the next decade—will drive discovery in astronomical research. Planetariums can and should be used for the advancement of scientific research. Current facilities such as the Hayden Planetarium in New York City, Adler Planetarium in Chicago, Morrison Planetarium in San Francisco, the Iziko Planetarium & Digital Dome Research Consortium in Cape Town, and Visualization Center C in Norrköping are already developing software which ingests catalogs of astronomical and multi-disciplinary data critical for exploration research primarily for the purpose of creating scientific storylines for the general public. We propose a transformative model whereby scientists become the audience and explorers in planetariums, utilizing software for their own investigative purposes. In this manner, research benefits from the authentic and unique experience of data immersion contained in an environment bathed in context and equipped for collaboration. ***Consequently, in this white paper we argue that over the next decade the research astronomy community should partner with planetariums to create visualization-based research opportunities for the field. Realizing this vision will require new investments in software and human capital.***

# RECOMMENDATIONS:
1) Planetariums need incentivization to open their facilities to local researchers. While the majority of the necessary infrastructure already exists, institutions will still incur operating costs, and some additional equipment may also need to be purchased. Grant opportunities should open for planetariums to be used as research visualization facilities by local academics, universities, observatories, etc. In a facility based model these planetariums could operate much like an observatory or supercomputing facility, with researchers applying for and purchasing time.
2) Hybrid positions between planetarium work and academic research should be enabled and/or prioritized. Joint positions can help bridge the cultural divide between the academic and planetarium communities.
3) Software development should be encouraged between open source research tools and open source planetarium tools, and the synergy between tools should be prioritized in funding opportunities.
4) Truely big data will require in-situ visualization. Close coordination between major research facilities and a planetarium visualization network is needed to enable exploration of these datasets.

5) Visualization in a planetarium requires adapted or even entirely new methods for constructing visual representations, rendering, and facilitating human-computer interaction. Joint research with experts in data visualization and human-computer interaction needs to be fostered.

# KEY ISSUES:

## Modern Planetariums as Immersive Data Visualization Facilities

***Issue #1: Researchers need to see planetariums as some of the most capable visualization locations on the planet***

Astronomy captures the public's imagination, so much so that it has its own dedicated public outreach venue: the planetarium. Worldwide, there are just over 4,000 planetariums that serve roughly 150 million visitors annually. The United States is the country with the highest number of planetariums at just over 1,600 (Petersen 2018). Over the last few decades, the planetarium community has largely transitioned from optomechanical to digital planetariums, allowing them to transition from displaying a static but beautiful simulation of the nighttime sky to visualizing the dynamic, three-dimensional universe in which we live.

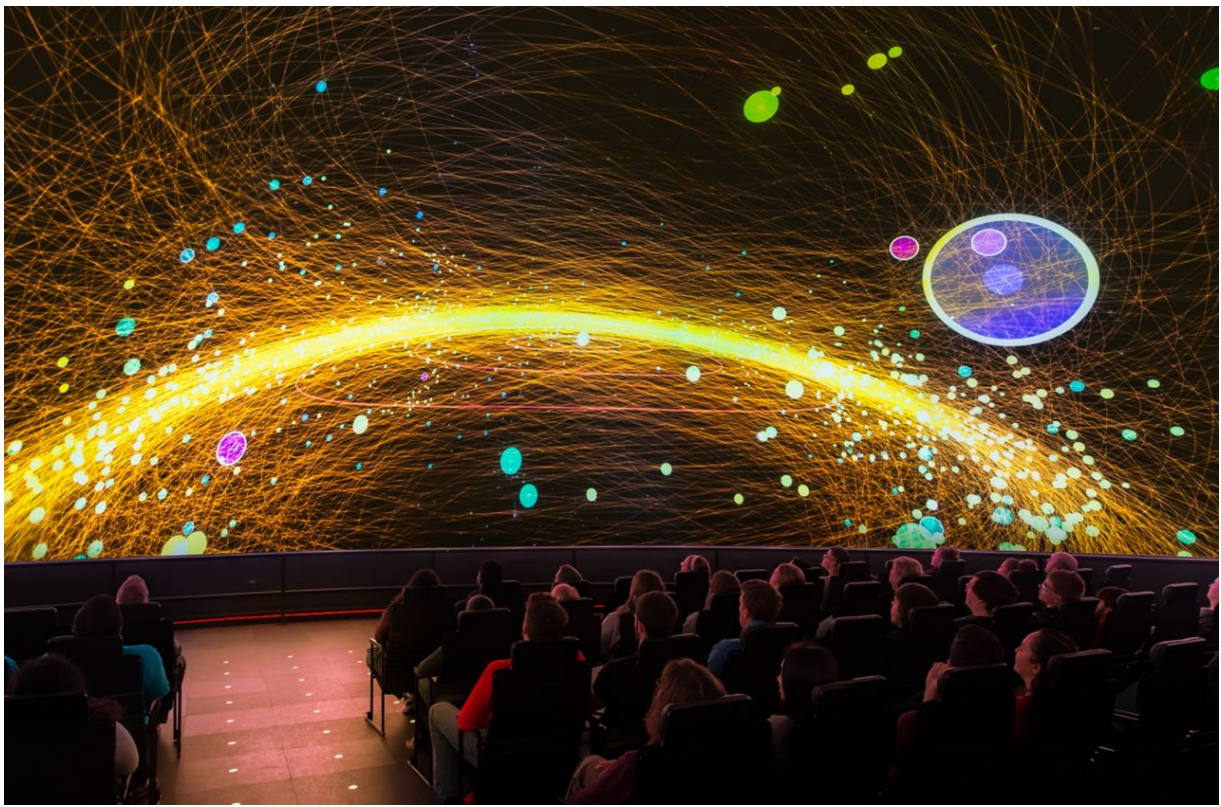

Figure 1. An audience at the Adler Planetarium in Chicago being immersed in a visualization of the Kuiper Belt.

Unfortunately, many academics have the notion of planetariums as old-school teaching machines of little value to scientists (Kwasnitschka 2017). This is despite the fact that modern digital projection domes are multimillion dollar facilities with state of the art hardware and are among the most capable visualization environments on the planet (see Figure 1 for an example of a modern day use of the planetarium for visualizing current data). Planetarium displays feature high resolutions (some as great as 80 Megapixels), high dynamic range, high frame rate (typically 60 frames per second), immersive (two pi steradians), and sometimes stereoscopic capabilities. Planetarium software comes with an extensive, dynamic, digital model of astronomical phenomena and can support large datasets (real-time rendering of one billion Gaia stars has been demonstrated). Many domes are large enough to enable collaborations of dozens or even hundreds to view a visualization simultaneously, and most can be networked, allowing multiple sites to synchronise views (a technology called "domecasting"). While many of these features exist on a desktop or VR device, the combination makes the planetarium a uniquely powerful visualization facility.

In van Wijk's 2005 analysis of "The Value of Visualization," high initial startup costs are blamed for why more scientists do not employ novel visualization methods in their research. Utilizing visualization environments that have already been established greatly reduces this barrier. We argue that promoting planetariums as dual use facilities makes much more economic sense than investing in expensive, bespoke, visualization displays for relatively few users.

## Planetariums as unique environments for scientific collaboration

***Issue #2: The astronomical era of big data demands visualization tools and facilities which allow researchers to collaborate while immersed in scientific context***

Exploration of large, complex data sets requires fast, precise, multidimensional data visualization as well as extremely broad expertise. The era of big astronomical data is not only "big" in the sense of having many elements, but the data we need to understand holistically can have attributes derived from many different kinds of observational and theoretical techniques, with complex biases, errors, and selection functions. Thus, having more than one pair of eyes on complex 3D explorations is critical to construct novel hypotheses efficiently. Domes provide this opportunity where other three dimensional visualization methods (e.g., VR) do not; being able to be in the same physical, visual space with colleagues during exploration is key to fast discovery. It is often in the raised eyebrow or emphatic gesture of a collaborator that one truly perceives skepticism or surprise. This need for 3D visualization is especially true in the age of Gaia, our brand new window onto the kiloparsec scale. The immersive experience of a large medium or even small sized dome is critical to unlocking the complexity of our galaxy, and the coming decade offers the first opportunity to embrace partnerships between academia and planetariums that will allow scientists to explore data together.

## Planetariums and Astronomical Big Data

On 25 April 2018, the European Space Agency's Gaia mission published its second data release, which contained parallaxes and proper motions for ~1.4 billion stars including ~8 million radial velocities (Gaia Collaboration et al. 2018; Lindegren et al. 2018). The size and scope of this foundational astrometric work brings to life the six-dimensional spatial and velocity structure of the galaxy and beyond with a magnitude 10,000 times above what was previously possible. Digesting the scope of Gaia DR2's structural information was an immediate desire for many astronomers and lent itself naturally to visualization facilities. Consequently, planetariums were at the forefront of Gaia DR2 data ingestion, modifying software packages to be able to render the scope of the data through both position and velocity space (Alsegard 2018 masters thesis). In June 2018, just two months after ESA released the Gaia DR2 catalog, the Flatiron Institute in New York hosted a "sprint" that brought together approximately 90 astronomers from across the globe to take one week dedicated to deep scientific investigations of the data. As part of the event, all astronomers were invited to an evening experience in the Hayden planetarium where they were immersed in the 6D scope of Gaia (both spatial and velocity) and select researchers presented their scientific investigations, findings, and questions through curated datasets prepared ahead of time. The experience led to unique and impressive questions which drove conversation and left many in the audience wanting an equal opportunity to see their own data displayed and investigated in a similar manner. Arguably, just one hour of an immersive dome experience of curated datasets led to a deeper understanding of Gaia DR2, new collaborations between researchers in the room, and the seeds for new discoveries that will be published in the coming years.

## Planetariums as Data Ingestion Engines

***Issue #3: Transferring orders of magnitude more data than previous decades requires novel methods for data ingestion and display***

The next decade will only bring more impressive orders of magnitude jumps in data ingestion and more opportunities for planetarium experiences for researchers. The Large Synoptic Survey Telescope (LSST) will collect 30 terabytes of data per night. The Square Kilometer Array (SKA), a radio telescope array, will dwarf even LSST, producing 160 terabytes of data every second! These numbers are daunting. Fortunately, the astronomical community has led the sciences in preparing itself to handle, process, and extract meaning from these data streams, through open data practices (Pepe et. al 2016) and open-source software systems such as WorldWide Telescope (now operated by the American Astronomical Society; Szalay and Gray 2001) and the Virtual Observatory—with aims to put all of the world's astronomical data (images, catalogs, even simulations) online in standardized and self-described data formats—and the Astronomy Visualization Metadata standard (Hurt et al. 2008), which embeds public outreach information directly into an astronomical image. For very large data sets, such as those that will be produced by SKA, it is impractical to transfer them to the planetarium. In cases such as these, a new hybrid solution is needed where the data is visualized locally at the data center and streamed to the planetarium. The academic astronomical community will benefit tremendously if we can merge research desires with advancements in planetarium capacity to ingest and visualize new datasets.

In 2013 the International Planetarium Society (IPS) established the "Data to Dome Initiative" (SubbaRao 2013). This effort was designed to prepare the planetarium community for the oncoming big data era in astronomy. In its first five years, this initiative has engaged stakeholders including planetariums, national observatories, planetarium vendors, and scientists (from corporate, government, museum, and academic communities). Significant progress has been made so far. This includes the establishment and implementation of new data standards for streaming content, numerous professional development efforts, and an active and engaged online community.

Digital planetariums are networked, allowing them to stream in live datasets. A new standard that emerged from the "Data to Dome" initiative is aptly called data2dome (Christensen et. al, 2016). This initiative makes it possible for content producers such as space agencies and observatories to publish new content directly to a planetarium operator's console. These data2dome feeds are now supported by several of the major planetarium software vendors, with full community support only a year or two away. Networking also enables remote collaboration through domecasting, which synchronizes content on planetariums in multiple locations. The Adler Planetarium uses domecasting in its Kavli Fulldome Lecture Series to simultaneously stream immersive public lectures to 15–20 planetariums around the world.

## Planetarium Software Infrastructure

There are roughly 10 planetarium software packages which dominate the market. These packages typically contain traditional planetarium functionality with the ability to fly through a digital 3D model of the universe. They allow the user to import their own datasets to extend that digital model as well. These software packages are feature rich, but also quite complex, and mastering their use takes significant effort (for example the user guides often exceed 1,000 pages). The current generation of planetarium software provides quite sophisticated data presentation functionality. But researchers attempting to use the software for data exploration will likely find these software packages lacking—they simply were not designed for these purposes. **We recommend a strategy of interoperability over integration. Following this strategy will allow researchers to interact with planetarium visualizations using the analysis environments they are already familiar with (for example, Jupyter notebooks or glue). The interoperability strategy also requires a significantly smaller development effort than adding data analysis capabilities to the extant planetarium software packages.**

Planetarium software packages consist of both commercial products and open source projects. Here we highlight two freely-available open source projects, alleviating any associated financial obstacle for academics: **AAS Worldwide Telescope** and **OpenSpace**. We believe that these software packages hold the most potential to engage researchers in accessing the planetarium over the next decade. In addition to being freely available and "dome ready" for all planetariums, these software packages already incorporate significant interoperability with research software.

**AAS WorldWide Telescope:** From its start as a Microsoft Research project, the AAS WorldWide Telescope (WWT) has always tried to serve both the outreach and research communities (Rosenfield et al. 2018). While WWT has enjoyed wide use in the planetarium community, it also supports data standards and interoperability protocols in general use in the astronomical community (e.g., FITS, VOTable, SAMP, AVM) and is considered by the IVOA to be a VO compliant tool[1] ([http://www.ivoa.net/astronomers/applications.html](http://www.ivoa.net/astronomers/applications.html)). Since the American Astronomical Society has taken over the stewardship of WWT, significant effort has been directed towards preparing the software for the future of research data visualization. The core effort has been to port the original Windows codebase (the version that historically runs in planetariums) to become a reusable JavaScript library founded on the WebGL rendering technology and a backing suite of web data services. Not only has this work led to the creation of a robust web application allowing users to explore the Universe from the comfort of their browser, it has enabled the WWT rendering engine to be embedded inside modern, web-based scientific analysis applications such as [Jupyter](), via the [pywwt]() Python module. Because pywwt can also control the Windows version of WWT using the same programming interface, a researcher or a student can prepare a visualization on their personal laptop and then seamlessly share it with colleagues in projection on a full planetarium dome. The same code can then serialize the visualization as a freestanding bundle of HTML, JavaScript, and data, for delivery as a customized Web interactive or an interactive figure in a scholarly publication.

**OpenSpace:** OpenSpace is an open source interactive data visualization software designed to visualize the entire known universe and to portray our ongoing efforts to investigate the cosmos. It was funded in 2016 by a NASA Science Mission Directorate education cooperative agreement notice (CAN). The development of this visualization tool is spearheaded by the American Museum of Natural History with strong partnerships at Linköping University, NASA's Community Coordinated Modeling Center, New York University, and the University of Utah. OpenSpace is a natural evolution of planetarium software, building heavily from data exploration tools such as SCISS's Uniview, which hosted the digital universe atlas curated by scientists at the Hayden Planetarium. The uniqueness of OpenSpace is in its versatility of exploring objects at the micro level (e.g., small topographic structures on the Earth, Moon, Mars, etc.) and fluidly switching to exploring objects at the macro level (e.g., large scale structure of the Universe mapped by the Sloan Digital Sky Survey galaxies or the cosmic microwave background). The software readily ingests datasets as they are made available (e.g., Kepler or TESS discovered planets, near-Earth asteroids, chemical abundance maps such as those provided by Apogee or Galah), and its open source code written primarily in C++ can be manipulated easily.

---

[1] [http://www.worldwidetelescope.org/webclient](http://www.worldwidetelescope.org/webclient)

# STRATEGIC PLAN:

## Partnered Software Development between Planetariums and Academia

Over the past decade, research scientists in astronomy have taken to open source code for analysis, and given the rich multi parameter nature of nearly all scientific investigations, multi-dimensional linked data exploration tools are highly desired. **Aladin,** for instance, is an interactive sky atlas tool allowing the user to visualize digitized astronomical images or full surveys, to superimpose entries from astronomical catalogues or databases, and to interactively access related data and information from the Simbad database, the VizieR service, and other archives for all known astronomical objects in the field. **Glue** as another example is a python library to explore relationships within and between related datasets. It has linked statistical graphics packages, full scripting capability, and robust flexibility at propagating selections across data sets.

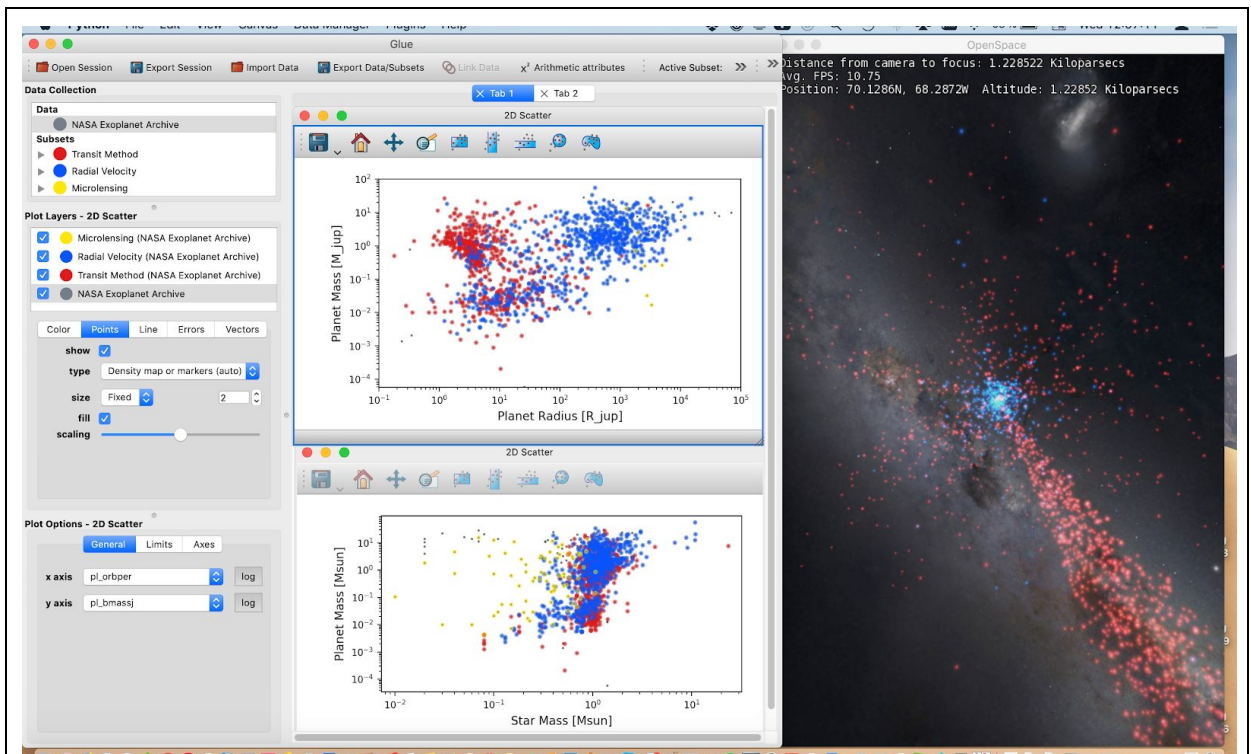

Figure 2. An example of the collision of planetarium software (OpenSpace at right) with astronomical research software (glue at left). In this case example glue makes a call to OpenSpace and can be used to fluidly change what appears in the planetarium as one edits datasets in standard glue plotting mechanisms.

The key to advancing the use of planetariums as scientific venues is cultivating interoperability between planetarium software and commonly-used research tools. As an example of what we recommend as a synergistic tool for the next decade, participants at a recent Dagstuhl seminar modified glue and OpenSpace so a user could fluidly move between the analysis tool to interactive data representation in a dome. Figure 2 shows what the interface looks like. In this example, glue controls the planetarium dome as a researcher analyzes data in a traditional way, with the results projected in the immersive environment of the dome. This data analysis works best for large astrometric data projects such as Gaia, where large scale structure in XYZ or velocity structure in UVW are critical to evaluating the science. Advancements are continuing in open source software development for astronomical research and they are increasingly prioritized by community leaders (e.g., white papers by D. Norman, "The Growing Importance of a Tech Savvy Astronomy and Astrophysics Workforce," A. Smith, "Elevating the Role of Software as a Product of the Research Enterprise," and E. Tollerud, "Sustaining Community-Driven Software for Astronomy in the 2020s"). ***We propose the next decade prioritize opportunities for partnerships between planetarium developed software (which enables advanced visualizations in existing infrastructure) with research focused software (which enables detailed scientific analysis of data).***

## Expand Opportunities for Researchers to Use Planetariums for Data Exploration

The United States has more planetariums than any other country in the world. While facilities are overwhelmingly used as informal education portals, the Hayden Planetarium at the American Museum of Natural History in New York City and Adler Planetarium in Chicago have established astrophysics research departments as well. The mission statements for both planetariums make very clear a strong desire for the facility to be used not just as an education resource, but as an astronomical resource as well.

Despite this, the best examples of researchers using planetariums for research purposes come from outside the United States. The Iziko Planetarium and Digital Dome (Marchetti & Jarrett 2018) in Cape Town opened in 2018 after a renovation. That renovation was supported in part by a consortium of local universities hoping to use the planetarium as a big data visualization facility for projects such as the multi-national Square Kilometer Array. The planetarium is closed to the public on Mondays when it is used exclusively by scientists, including researchers from disciplines outside of astronomy. The Iziko also holds regular 'Data to Dome' public events, where these scientists directly share their science with the public. This is a great example of the kind of positive feedback loop that can occur by inviting scientists into the dome.
***In the coming decade our strategic plan would encompass an increase in the number of academics working in planetariums and replication of the advancements made by our global partners working between research and education.***

## Encourage Partnerships Between Universities or Research Centers and Local or Campus Planetariums

In the United States there are a few planetariums co-located with major centers of astronomical research. These planetariums will play a key role in realizing the vision of planetariums as research facilities. The 'Imiloa Astronomy Center of Hawaii in Hilo was designed to be a comprehensive educational facility that would showcase the connections between the rich traditions of Hawaiian culture and the groundbreaking astronomical research conducted at the summit of Mauna Kea. Located next to the base facilities of many of the Mauna Kea observatories, provides many resident and visiting astronomers easy access to 'Imiloa. Similarly, the Flandrau Planetarium on the University of Arizona campus in Tucson is located across the street from both the NOAO headquarters and the LSST headquarters, and therefore should be able to showcase exciting scientific discoveries as they are realized from these research facilities.

There are many planetariums located on university campuses. Frequently, these planetariums include a mandate to support research in disciplines other than astronomy as well. Some of the major university-based planetariums in the USA include: Fiske Planetarium at the University of Colorado Boulder, the Digital Visualization Theater at the University of Notre Dame, the Ho Tung Visualization Laboratory at Colgate University, Abrams Planetarium at Michigan State University, Jordan Planetarium at the University of Maine, and Charles W. Brown Planetarium at Ball State University.

While much software development is happening at the largest planetariums (e.g., Hayden, Adler, Morrison), this needs to be disseminated across the 1,600 planetariums in the U.S. Most importantly, campus planetariums or those close to universities or research facilities should be incentivized to utilize the software developments made by the largest planetariums and partner with local academics on advancing their research.

**Acknowledgements**
This white paper was started at Dagstuhl Seminar 19262: *Astrographics: Interactive Data-Driven Journeys through Space*.